\documentclass[11pt,a4paper]{article}

\usepackage[margin=2.2cm]{geometry}
\usepackage[T1]{fontenc}
\usepackage[utf8]{inputenc}
\usepackage{lmodern}
\usepackage{microtype}
\usepackage{amsmath,amssymb,bm}
\usepackage{booktabs,multirow,array,tabularx}
\usepackage{graphicx}
\usepackage{xcolor}
\usepackage{siunitx}
\usepackage{enumitem}
\usepackage{hyperref}
\usepackage{cleveref}
\usepackage{authblk}
\usepackage{caption}
\usepackage{subcaption}
\usepackage{placeins}
\usepackage{fancyhdr}
\usepackage{url}

\hypersetup{colorlinks=true,linkcolor=blue!45!black,citecolor=blue!45!black,urlcolor=blue!45!black}
\title{\textbf{Towards AI-Agent-Accelerated Research in High-Energy Physics:
A Case Study in Monte Carlo Analysis}}
\author[1]{Tianqi Gao}
\author[2]{Ruobing Jiang\thanks{Corresponding author: \texttt{ruobing@stu.pku.edu.cn}}}
\author[2]{Dawei Fu}
\author[2]{ Andrew Michael Levin}
\author[2]{Qiang Li}
\author[1]{Matthew Kenzie}
\affil[1]{Cavendish Laboratory, University of Cambridge, Cambridge CB3 0HE, United Kingdom}
\affil[2]{State Key Laboratory of Nuclear Physics and Technology, School of Physics, Peking University, Beijing 100871, China}
\date{}

\begin{document}
\maketitle

\begin{abstract}
We investigate how far contemporary LLM-based agents can participate in multistep computational high-energy-physics research under reproducible human supervision. As a realistic case study, we use a previously published CEPC analysis of Bell observables in
$e^+e^-\to ZH$, $Z\to q\bar q$, $H\to ZZ^*\to4\mu$. The reproduction is instruction-constrained rather than formally blinded: the agent is explicitly instructed not to consult or use the published numerical results as analysis targets during implementation, although no technical information-access barrier is imposed. The analysis definition is fixed before the formal benchmark  comparison. The agent participates in workflow planning, code and configuration inspection, Monte Carlo production, anomaly diagnosis, targeted-control construction, statistical analysis, and result organization, while state-changing actions and final interpretation remain under human control. The resulting detector-level $I_3$ central values reproduce the four published benchmark points to within approximately 2\%, without retuning the detector or analysis configuration to the published values. The agent is then used for an open-ended extension for which no target answer is supplied. Using 200,000 fast-simulated events, the stage-resolved audit identifies muon isolation as the largest single contribution to the hard-to-detector shift. The subsequent diagnostic sequence connects the effect to angle-dependent isolation efficiency, proximity to the hard $Z\to q\bar q$ system, and mixed electromagnetic activity from muon final-state radiation and hadron decays. This case study shows that the useful role of an AI agent in HEP is not only faster code generation, but faster scientific iteration: maintaining complex software state, testing alternative hypotheses, constructing controls, and propagating validated results through an analysis. The study motivates systematic benchmarking of scientific agents across historical HEP analyses, with correctness, human intervention, and researcher time efficiency as explicit evaluation metrics.
\end{abstract}

\section{Introduction}

A substantial fraction of computational high-energy-physics (HEP) research effort is spent not on the final physics interpretation itself, but on constructing and maintaining long analysis chains: configuring event generation and detector simulation, reproducing selections, tracing software and convention mismatches, debugging distributed production, validating intermediate objects, evaluating statistical procedures, and constructing repeated control studies. Such work can occupy significant fractions of PhD and postdoctoral appointments and can limit the number of scientific questions that an individual researcher or a small group is able to pursue. Large-language-model (LLM) systems raise a broader possibility than simply writing code faster: scientific agents may allow researchers to supervise several computational investigations, delegating part of the implementation, diagnosis, and validation cycle while retaining control over the scientific objective and final interpretation. The relevant question is therefore not only whether AI can accelerate isolated coding tasks, but whether it can increase the number of validated scientific questions that can be advanced per unit of human researcher time.

This possibility requires a distinction between AI-assisted programming and agentic scientific research. In a conventional AI-assisted workflow, a model may write a function, explain an error message, or suggest a plotting routine while the scientific strategy and sequence of actions remain external to the model. An agentic workflow poses a stronger question: can an AI system inspect heterogeneous scientific evidence, construct and execute a multistep computational plan, diagnose failures or inconsistencies, formulate testable hypotheses, design targeted controls, use scientific software tools, and maintain the state of an analysis across repeated iterations? For computational particle physics, where a single result may involve matrix-element generation, parton showering, detector simulation, reconstruction, statistical inference, batch computing, and multiple validation layers, this provides a demanding test of scientific-agent capability. Before agentic workflows can be considered a reliable way to increase HEP research productivity, two questions must therefore be addressed: how much of a realistic analysis can a contemporary agent actually carry out, and under what controls can its output be considered scientifically trustworthy?

Recent work has begun to evaluate LLM-based agents on increasingly realistic scientific tasks. ScienceAgentBench assesses agents on data-driven research problems extracted from peer-reviewed studies, while Agent Laboratory explores LLM-agent workflows spanning literature review, experimentation, and report generation~\cite{Chen2025ScienceAgentBench,Schmidgall2025AgentLaboratory}. More recent benchmark efforts have emphasized controlled and reproducible evaluation across broader scientific workflows~\cite{Bragg2026AstaBench}. Within HEP, evidence-grounded and agent-assisted systems have also been developed for scientific literature retrieval and question answering in muon collider research and silicon-detector R\&D~\cite{Jiang2026MuonRAG,Gao2026SiliconRAG}. These studies illustrate that scientific agents can already contribute to specific components of the research process, but they do not yet establish how reliably an agent can reconstruct and extend a complete collider-analysis workflow. The present work addresses this complementary question by testing whether an agent can execute and extend a complete collider-analysis workflow under explicit provenance and human-control constraints.

We investigate these questions using a previously published collider-phenomenology analysis as a realistic case study. We choose the CEPC study of Bell inequalities and quantum entanglement in Ref.~\cite{WuEtAl2025CEPC}, which studied Higgs production at a future electron--positron collider using Monte Carlo simulation and angular quantum-state tomography. The physical process is
\begin{equation}
 e^+e^- \to ZH,\qquad
 Z\to q\bar q,\qquad
 H\to ZZ^*\to\mu^+\mu^-\mu^+\mu^- ,
\end{equation}
at $\sqrt{s}=250$~GeV. Quantum-information observables extracted from collider angular distributions provide a useful benchmark because they depend on the complete analysis chain rather than on a single inclusive rate. For a spin-1 pair, Bell tests can be formulated using the Collins--Gisin--Linden--Massar--Popescu (CGLMP) inequality~\cite{CGLMP2002}, with the relevant density-matrix information reconstructed from the decay angles of the two $Z$ bosons. Related spin-density-matrix, entanglement, and Bell observables have been studied in top-quark, diboson, Higgs, and tau systems~\cite{Barr2022,AguilarSaavedra2023HZZ,AshbyPickering2023,Barr2024Review}, while measurements at the LHC have established that collider quantum observables are experimentally accessible~\cite{CMS2024Entanglement,ATLAS2026ZZEntanglement,ATLAS2024Top}. The reproduction performed here is \emph{instruction-constrained} rather than formally blinded: the agent was explicitly instructed not to consult or use the published numerical results as analysis targets during implementation, but no technical information-access barrier was imposed. The analysis definitions and production configuration were fixed before the formal comparison with the published numerical values, and no detector parameter or selection was subsequently retuned to improve the agreement.

Reproduction of the known observable constitutes only the first capability test. After the benchmark analysis is established, the agent is asked to pursue an open-ended extension for which no target numerical answer, dominant reconstruction stage, topology variable, or physical mechanism is supplied in advance. The task is to determine whether the detector and reconstructed-object definitions modify the Bell observable and, if so, to identify the origin of the distortion. This is nontrivial because an angular observable is affected not only by four-vector resolution: an angle-dependent reconstruction or selection efficiency can reweight the underlying decay-angle distribution and thereby sculpt the reconstructed density matrix. Using a 200,000-event fast-simulation production, the workflow constructs complementary fixed-vector and stage-vector audits, identifies muon isolation as the largest single contribution to the hard-to-detector change in the CGLMP estimator $I_3$, and subsequently traces the effect to non-uniform angular efficiency, proximity to the hard $Z\to q\bar q$ system, and mixed electromagnetic activity from muon final-state radiation and hadron decays. These intermediate targets were not supplied as predefined answers; they emerged through successive hypotheses and deterministic controls applied to the frozen production.

Taken together, the reproduction and open-ended extension provide complementary tests of execution and scientific diagnosis. Throughout both components, the agent participates in workflow planning, source-code and configuration inspection, sample production, anomaly diagnosis, diagnostic-script construction, statistical analysis, and result organization. Its exploratory role is separated from the evidential chain used for the final result: frozen definitions, fixed random seeds, checksum-pinned artifacts, isolated diagnostic branches, deterministic scripts, and explicit human authorization for state-changing actions provide provenance barriers between exploration and validated production. The broader objective is to understand whether such systems can reduce technical overhead and shorten the cycle between a scientific question, a computational test, and a validated answer, thereby increasing the research capacity of individual physicists, PhD students, postdoctoral researchers, and small groups. A single case study cannot establish that general capability, but it provides a basis for future systematic benchmarks across historical HEP analyses in which numerical accuracy, failure recovery, human intervention, computational cost, and active researcher time can be measured explicitly.

The remainder of the paper is organized as follows. Section~\ref{sec:agentic} describes the agent environment, initial research instructions, human intervention, and provenance and validation protocol. Section~\ref{sec:case} specifies the reference HEP task, including the Bell observable, simulation chain, reconstruction, and frozen analysis selections. Section~\ref{sec:benchmark} describes execution of the benchmark analysis and the quantitative comparison with the published result. Section~\ref{sec:sculpting} presents the open-ended stage-resolved diagnosis, and Sec.~\ref{sec:mechanism} follows the diagnostic sequence to the angular, topological, and generator-level interpretation of the isolation effect. Section~\ref{sec:discussion} discusses reliability, human--agent division of labor, research productivity, and limitations. Section~\ref{sec:conclusion} summarizes the case study and outlines a broader program for benchmarking AI agents across HEP research tasks.

\section{Agent setup and research process}
\label{sec:agentic}

\subsection{Agent environment and initial research instructions}

The computational environment used Python~3.9.25, ROOT~6.40.02, and HTCondor~24.12.16 for analysis, validation, and distributed production. OpenAI ChatGPT and Codex were used for interactive scientific reasoning, code and configuration inspection, workflow planning, script generation, execution support, diagnostic design, statistical analysis, and result organization. Anthropic Claude was used for language editing and manuscript consistency checks rather than for production of the numerical physics results. Exact software versions, command lines, random seeds, configuration files, generated artifacts, and validation records were retained in the project provenance. The study does not attempt a model-to-model comparison of language-model performance; the object of evaluation is the scientific workflow and the research tasks completed within it rather than a single immutable model release.

The project began from a staged human instruction rather than from a complete analysis script. The initial task was deliberately restricted to an audit of the available local material before any event generation or detector production was attempted. Reconstructed from the project record, the initial instructions required the agent to read the project guidance in \texttt{AGENTS.md} and inspect the locally available paper and \LaTeX{} sources; reconstruct the analysis requirements and trace physics choices to explicit source locations where possible; record unspecified quantities as ambiguities rather than infer them silently; prepare the analysis-specification and reproducibility records; and stop before Monte Carlo production, software installation, or other large state-changing operations. Subsequent instructions progressively authorized theory implementation, lightweight unit and closure tests, generation planning, development samples, and finally controlled enlarged production. The project rules also prohibited silent physics assumptions and ad hoc tuning: estimator normalization and conventions had to be established through first-principles closure tests rather than empirical rescaling to match a known answer.

The reproduction was \emph{instruction-constrained} rather than formally blinded. The agent was explicitly instructed not to consult or use the published numerical results as analysis targets during implementation, but no technical information-access barrier was imposed. We therefore do not claim information-access blinding. The analysis definitions and production configuration were fixed before the formal benchmark comparison, and no detector parameter or selection was subsequently retuned to improve agreement with the published result.

\begin{figure}[tbp]
\centering
\includegraphics[width=0.92\linewidth]{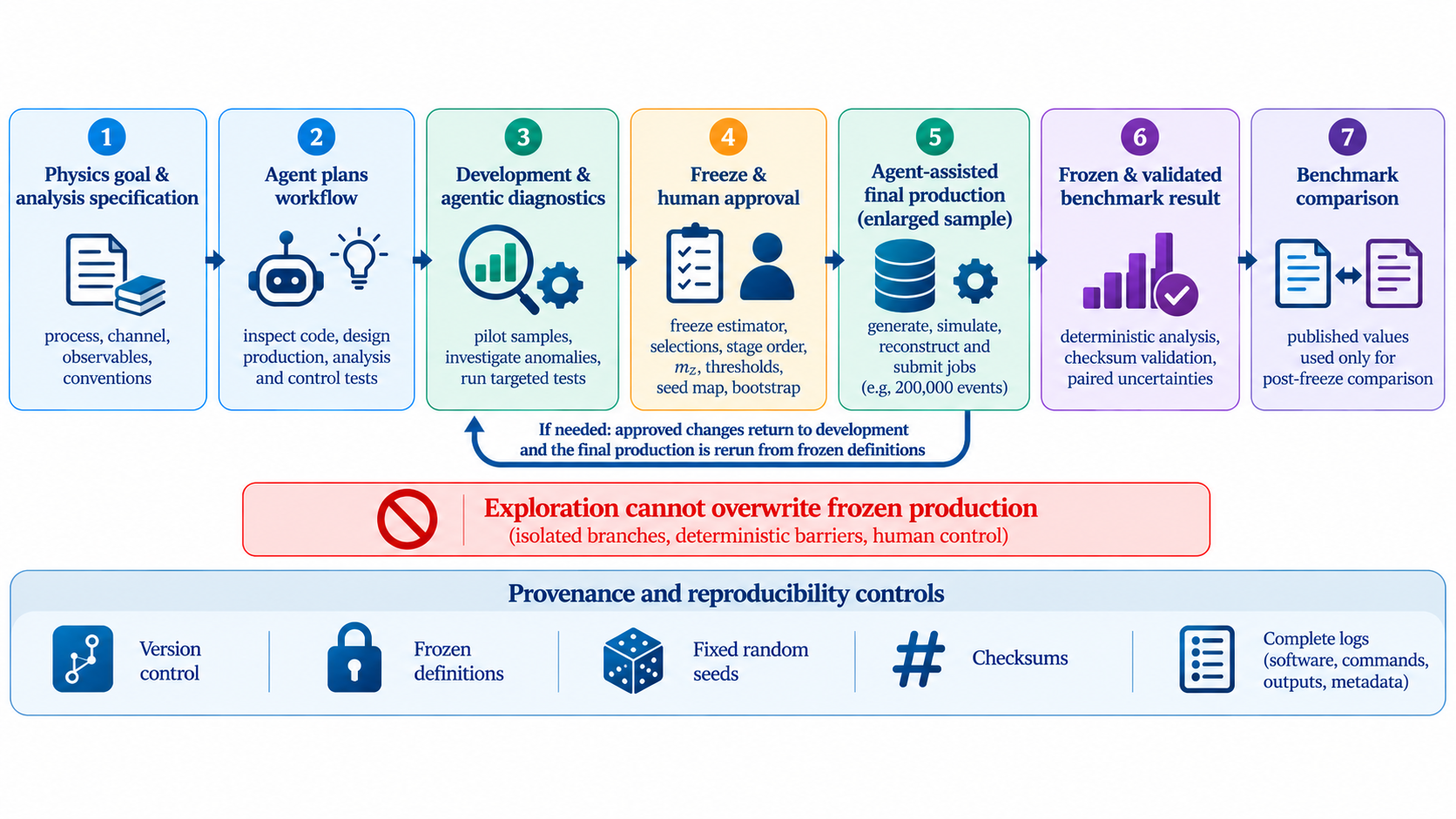}
\caption{
Schematic of the provenance-controlled agentic workflow used in this study. A human researcher first defines the scientific objective, required physics conventions, and permitted scope of the analysis. The agent then audits the available evidence, plans and executes development tasks, inspects intermediate outputs, and proposes diagnostic controls. Before the enlarged production is evaluated, the estimator, selections, stage definitions, random seeds, and bootstrap procedure are frozen under human approval. The resulting frozen benchmark output is validated deterministically and only then compared formally with the published CEPC values. A subsequent read-only agentic investigation uses the frozen production for stage-resolved diagnosis, mechanism studies, and robustness checks without modifying validated artifacts. Version control, fixed seeds, checksums, complete logs, deterministic validation, and explicit human authorization provide the provenance boundary between exploratory reasoning and final physics results.
}
\label{fig:agentic}
\end{figure}

\subsection{Human intervention and division of responsibility}

The workflow was not intended as a test of unsupervised autonomy. Instead, it was designed to determine how much of a realistic computational HEP analysis could be delegated to an agent while preserving human control over the scientific objective, state-changing operations, and final interpretation. The human researcher defined the initial physics task, supplied the required conventions and project constraints, authorized production submissions and modifications to frozen definitions, assessed whether proposed controls were physically meaningful, and retained responsibility for the final scientific claims. The agent carried out much of the intermediate work: inspecting software and configurations, reconstructing analysis requirements from source material, writing and revising scripts, connecting information across the generator, shower, detector, reconstruction, and statistical layers, diagnosing software and physics mismatches, proposing competing explanations, implementing controls, organizing outputs, and checking numerical consistency.

\begin{table}[tbp]
\centering
\small
\caption{Representative division of responsibility between the scientific agent and the human researcher. The table records major intervention points rather than every interactive message.}
\label{tab:humanroles}
\begin{tabularx}{\linewidth}{
>{\raggedright\arraybackslash}X
>{\raggedright\arraybackslash}X
>{\raggedright\arraybackslash}X
>{\raggedright\arraybackslash}X}
\toprule
Research stage & Agent role & Human intervention & Purpose \\
\midrule

Initial audit and benchmark constraint
&
Inspect local paper/source material, reconstruct the required workflow, and record unresolved ambiguities without using the published numerical $I_3$ values as tuning targets
&
Define the physics objective, source material, tool restrictions, and instruction-constrained benchmark protocol
&
Prevent premature implementation, silent assumptions, and tuning to the published result
\\

Theory implementation
&
Implement conventions and construct closure and consistency tests
&
Require first-principles validation and reject ad hoc normalization or tuning
&
Establish the estimator before detector progression
\\

Production setup
&
Construct and test generation, showering, detector-simulation, extraction, and batch workflows
&
Authorize state-changing production submissions and major configuration changes
&
Maintain control over the nominal production state
\\

Development anomaly and mechanism study
&
Propose isolation-focused controls, test topology variables, reject unsuitable proxies, and construct ancestry diagnostics
&
Approve frozen stage definitions and assess the physical meaning of proposed controls
&
Prevent post-observation retuning and retain human responsibility for interpretation
\\

Manuscript preparation
&
Organize numerical results, draft descriptions, and check internal consistency
&
Select claims, revise interpretation, and approve the submitted text
&
Preserve author responsibility for the publication
\\

\bottomrule
\end{tabularx}
\end{table}

The project did not prospectively record active human time in a form suitable for a controlled productivity measurement. We therefore make no numerical claim for a speed-up factor. Human-intervention time, total wall-clock time, computational cost, and the number of validated tasks completed per researcher-hour are instead proposed as explicit metrics for future scientific-agent benchmarks.

\subsection{Instruction-constrained benchmark and provenance controls}

The benchmark tested whether the agent could reconstruct and execute an established collider-phenomenology calculation without using the published numerical result as an optimization target. Before the enlarged production result was evaluated, the observable definition, same-flavor pairing convention, stage order, four $m_{Z_2}$ thresholds, seed map, and bootstrap procedure were frozen. The formal comparison with Ref.~\cite{WuEtAl2025CEPC} was then performed as a post hoc validation, and no detector parameter, estimator convention, or analysis selection was subsequently modified in order to improve the numerical agreement. The enlarged sample retains the original development seeds and is therefore not statistically independent of the development sample; it should consequently be interpreted as a procedural confirmation of a frozen analysis rather than as an independent-data replication.

The workflow maintained a strict separation between exploratory reasoning and frozen production. Diagnostic branches could add output-only information, test alternative hypotheses, or execute control configurations, but validated production files were not implicitly replaced. File hashes, random seeds, software versions, command lines, resolved configurations, host information, timestamps, and completion records were persisted with the outputs. Numerical conclusions entered the analysis through deterministic scripts acting on pinned inputs, while immutable event identities and checksums were used to verify consistency between stages. Agentic reasoning could propose and implement diagnostics, but any state-changing modification to the nominal production or frozen analysis required explicit human authorization. Reliability therefore derives from the combination of agentic exploration, deterministic validation, provenance, and human scientific judgment rather than from treating the language model itself as a numerical authority.

\subsection{Representative agentic research episodes}

The practical scientific role of the agent is illustrated by the episodes in Table~\ref{tab:agentcases}. Each follows the same general pattern:
\[
\text{observation}
\rightarrow
\text{hypothesis}
\rightarrow
\text{targeted control}
\rightarrow
\text{deterministic outcome}
\rightarrow
\text{human interpretation}.
\]
These cases provide direct evidence of work beyond code generation: the agent repeatedly integrated information from source code, detector cards, logs, event-level outputs, and statistical results to propose and test explanations for unexpected behavior.

\begin{table}[tbp]
\centering
\small
\caption{Representative agentic diagnostic episodes. Physics conclusions were accepted only after deterministic controls and, where relevant, human approval of changes to the nominal workflow.}
\label{tab:agentcases}
\begin{tabularx}{\linewidth}{
>{\raggedright\arraybackslash}X
>{\raggedright\arraybackslash}X
>{\raggedright\arraybackslash}X
>{\raggedright\arraybackslash}X}
\toprule
Observed issue & Agent-proposed hypothesis & Control & Outcome \\
\midrule

Delphes card syntax around segmentation
&
A syntax-version mismatch, rather than a physics change, caused parser incompatibility
&
Translate only the incompatible syntax and compare the same-seed detector response
&
Event-level response retained; no physics parameter was tuned
\\

Muon-mass import changed dilepton masses
&
PYTHIA's LHE lepton-mass mapping redistributed recoil
&
Compare import conventions and fixed hard-vector closure
&
Source of the migration identified and quantified; hard-reference convention frozen
\\

Large isolation shift in the 1000-event development sample
&
Muon isolation was a candidate sculpting stage
&
Isolation bypass/loose controls, followed by a 200k frozen-definition audit
&
Development effect reduced under the final definition; 200k paired bootstrap provided the quantitative result
\\

HTCondor staging/publishing failures
&
Transport paths, rather than physics code, caused failed completion records
&
Isolated staging tests, checksums, and immutable-output checks
&
Recovery without rerunning validated physics products
\\

\bottomrule
\end{tabularx}
\end{table}

The later physical-mechanism study provides a further example of open-ended agentic diagnosis. Unlike the reproduction benchmark, this part of the project had no published target answer specifying the dominant reconstruction stage, the appropriate topology variable, or the physical composition of the isolation activity. The agent inspected the frozen production, formulated candidate explanations, and constructed controls to distinguish among them. One initially plausible interpretation used reconstructed jets to characterize nearby hadronic activity, but explicit validation rejected this proxy: the pre-overlap jet collection contains the signal muon itself, while the final cleaned-jet collection is defined downstream of the isolation decision. The analysis consequently moved to the two hard quarks from the associated $Z\to q\bar q$ decay as an upstream topology reference, followed by generator-ancestry diagnostics of the electromagnetic cone activity.

The benchmark reproduction and open-ended diagnostic study therefore probe different capabilities. The first asks whether an agent can construct and execute an established HEP calculation under explicit instructions not to optimize against the published answer. The second asks whether the same workflow can progress from an unexplained numerical effect to discriminating controls and a physically interpretable mechanism when no target explanation is supplied in advance. In both cases, the agent expands the amount of implementation and diagnostic work that can be delegated while leaving the evidential threshold for the final physics result unchanged.

\section{Reference HEP task and frozen analysis specification}
\label{sec:case}

This section specifies the HEP task against which the agentic workflow is evaluated. The purpose is not to introduce a new Bell-observable formalism, but to record the physics definitions, simulation configuration, reconstruction sequence, and selection conventions that the agent had to reconstruct, implement, and propagate through the Monte Carlo chain. These definitions were subsequently frozen before the formal benchmark comparison and were reused for the open-ended diagnostic studies.

\subsection{Bell observable and reference definition}
\label{sec:observable}

\subsubsection{Angular moments and \texorpdfstring{$I_3$}{I3}}

The analysis follows the angular convention and same-flavor pairing used in the reference CEPC study~\cite{WuEtAl2025CEPC}, based on the formalism of Ref.~\cite{AguilarSaavedra2023HZZ}. For each valid $4\mu$ configuration, the two opposite-sign dimuon candidates are labeled $Z_1$ and $Z_2$, with the larger and smaller dimuon masses, respectively. Reconstructing this convention correctly is part of the benchmark task because the Bell estimator is sensitive to the pairing and angular definitions.

The decay angles are defined in an orthonormal basis $(\hat r,\hat n,\hat k)$ constructed in the Higgs rest frame, reconstructed from the four-muon system. The $z$ axis $\hat k$ is the direction of $Z_1$ in this frame. With $\hat p$ the unit vector along the $e^-$ beam and $\cos\Theta=\hat k\cdot\hat p$, the remaining basis vectors are
\begin{equation}
 \hat r=\frac{\mathrm{sign}(\cos\Theta)}{\sin\Theta}
 \left(\hat p-\cos\Theta\,\hat k\right),
 \qquad
 \hat n=\hat k\times\hat r .
\label{eq:basis}
\end{equation}
Each muon is boosted from the Higgs rest frame into the rest frame of its parent $Z_i$. The solid angle $\Omega_i=(\theta_i,\phi_i)$ gives the direction of the $\mu^-$ from $Z_i$ in that frame, with $\theta_i$ measured from $\hat k$ and $\phi_i$ measured from $\hat r$ in the $\hat r$--$\hat n$ plane. The same basis is used for $Z_1$ and $Z_2$.

The spin-correlation coefficients are obtained as event averages of products of spherical harmonics $Y_{\ell m}(\Omega)$. With
\begin{equation}
 B_2=\sqrt{\frac{2\pi}{5}},
\end{equation}
the two correlation coefficients entering the benchmark are
\begin{align}
 C_{2,1,2,-1} &=
 \frac{(4\pi)^2}{B_2^2}
 \left\langle
 \mathrm{Re}\left[
 Y_{21}(\Omega_1)Y_{2,-1}(\Omega_2)
 \right]
 \right\rangle,
 \\
 C_{2,2,2,-2} &=
 \frac{(4\pi)^2}{B_2^2}
 \left\langle
 \mathrm{Re}\left[
 Y_{22}(\Omega_1)Y_{2,-2}(\Omega_2)
 \right]
 \right\rangle.
\end{align}
Following Ref.~\cite{WuEtAl2025CEPC}, the single-$Z$ coefficient $A_{20}$ is expressed through the second correlation coefficient as
\begin{equation}
 A_{20}=\sqrt{2}\left(C_{2,2,2,-2}-1\right).
\end{equation}
The CGLMP estimator~\cite{CGLMP2002,AguilarSaavedra2023HZZ} used throughout the benchmark and subsequent diagnostic studies is
\begin{align}
 I_3 = \frac{1}{36}\Big[&
 18+16\sqrt{3}
 -\sqrt{2}\,(9-8\sqrt{3})A_{20}
 \nonumber\\
 &-8(3+2\sqrt{3})C_{2,1,2,-1}
 +6C_{2,2,2,-2}\Big].
\label{eq:i3}
\end{align}
The explicit expression is retained because normalization and convention choices were among the quantities that had to be established and validated during development. Reference~\cite{WuEtAl2025CEPC} quotes numerical values of $I_3$ and of the correlation coefficients but not of $A_{20}$, so no standalone $A_{20}$ benchmark comparison is performed.

\subsubsection{Hard reference and event weights}

The frozen analysis specification defines the original LHE hard-muon four-vectors, before showering and detector simulation, as the immutable reference used in the fixed-vector audit. The pairing and angular construction are applied to these vectors once, after which the resulting angular kernels are retained while successive detector-stage masks are varied. We therefore use the term \emph{hard reference} rather than \emph{Born-level truth}: the generated event contains an explicit resonance history, while the later detector-level chain additionally includes showering, QED final-state radiation, detector response, and reconstruction.

Accordingly, the hard-to-detector difference is not interpreted as pure detector resolution. It can contain QED radiation, pairing and mass-window migrations, reconstructed four-vector response, and selection acceptance. The fixed-vector audit isolates the population-selection component by holding the angular kernels fixed, while the complementary stage-vector construction restores the reconstructed four-vectors available at each stage.

The MadGraph samples are unweighted leading-order samples. Stored LHE weights are nevertheless propagated through all estimators and bootstrap replicas so that the statistical procedure remains well defined for the general weighted case; for uniform weights this is algebraically equivalent to an unweighted mean.

\subsection{Simulation and event selection}
\label{sec:simulation}

\subsubsection{Event generation and showering}

The reference process is
\begin{equation}
 e^+e^- \to ZH,\qquad
 Z\to q\bar q,\qquad
 H\to ZZ^*\to\mu^+\mu^-\mu^+\mu^-,
\end{equation}
generated at $\sqrt{s}=250$~GeV with the resonance decay chain encoded at matrix-element level. The final production contains 200 random seeds with 1000 events per seed, giving 200,000 generated events. The original 20 development seeds are retained and 180 additional seeds enlarge the production. No additional branching-fraction factor is applied after generation because the decay-chain branching fractions are already contained in the generated process definition.

Matrix-element generation uses MadGraph5\_aMC@NLO version~3.7.1~\cite{Alwall2014}, followed by showering and hadronization with PYTHIA~8.317~\cite{Pythia83}. The frozen production steering has initial-state radiation disabled, QED final-state radiation and hadronization enabled, no pileup, and no beamstrahlung model. Beam polarization is inherited from the LHE input and is zero for this production. These settings are stated explicitly because part of the agentic task was to distinguish documented analysis choices from assumptions that were not specified in the reference material.

One development-stage control identified a convention dependence associated with PYTHIA's treatment of the massless LHE muons: recoil redistribution can shift dilepton masses by up to approximately $0.45$~GeV in a 1000-event control sample. This issue and the corresponding validation are described in Appendix~\ref{app:controls}. The frozen hard reference always uses the original LHE four-vectors.

\subsubsection{Fast detector simulation}

Detector response is simulated with Delphes~3.5.1~\cite{Delphes2014} using the CEPC detector card associated with Ref.~\cite{WuEtAl2025CEPC}. The card was translated only where necessary for compatibility with the installed Delphes syntax. That translation was treated as a software-compatibility intervention rather than as an opportunity to tune the detector model: same-seed validation was used to verify the response, and no detector parameter was adjusted to improve agreement with the published $I_3$ values. The diagnostic configuration writes additional intermediate branches while leaving the nominal physics modules and settings unchanged.

For the muon chain, the card has zero efficiency for $E<2$~GeV or $|\eta|>3$ and $99\%$ efficiency otherwise. Muon isolation operates on pre-isolation candidates from the muon-efficiency module and on particle-flow inputs from the filtered EFlow collection. The isolation cone has $\Delta R=0.5$, includes constituents with $p_T\ge0.5$~GeV, and requires a relative scalar-$p_T$ sum not exceeding $0.25$. Identified electrons and muons are removed from the filtered isolation input, so a signal muon does not directly contribute its own momentum to the isolation sum. No $\rho$ input is supplied, and the corresponding correction is therefore zero.

The reconstructed electron collection used by the electron veto follows the same nominal CEPC energy and acceptance requirement, with $E>2$~GeV and $|\eta|<3$ in the active region. Jets are clustered using FastJet~3.5.1~\cite{Cacciari2012FastJet} with the anti-$k_T$ algorithm~\cite{Cacciari2008AntiKt}, radius parameter $R=0.5$, and a transverse-momentum threshold of 20~GeV.

\subsubsection{Reconstruction and frozen analysis selection}

Rather than collapsing detector reconstruction into a single selection, the frozen task records the ordered reconstruction stages explicitly:
\begin{enumerate}[label=(\roman*)]
 \item hard reference;
 \item tracking geometry;
 \item tracking smearing;
 \item muon energy and pseudorapidity acceptance;
 \item stochastic muon efficiency;
 \item muon isolation;
 \item overlap removal;
 \item exactly four signal-matched reconstructed muons;
 \item exactly two jets together with the electron veto;
 \item opposite-sign pairing and the $Z_1$ mass window;
 \item nominal detector-level kinematics and selection.
\end{enumerate}
This ordering subsequently provides the basis for the open-ended stage-resolved diagnosis.

The same-flavor pairing follows Ref.~\cite{WuEtAl2025CEPC}: the larger dimuon mass is assigned to $Z_1$ and the smaller to $Z_2$. The strict on-shell requirement is
\begin{equation}
 \left|m_{Z_1}-91.1876~\mathrm{GeV}\right|<10~\mathrm{GeV}.
\end{equation}
The final ``nominal detector'' stage recomputes the dimuon pairing, rest-frame angles, and mass selections using the final reconstructed muon four-vectors. Its event count can therefore differ slightly from the preceding frozen-mask stage because of kinematic migrations. Four nested $Z_2$ requirements are evaluated:
\[
m_{Z_2}>0,\ 10,\ 20,\ 30~\mathrm{GeV}.
\]
Because these selections are nested, their statistical results are strongly correlated and are not interpreted as four independent confirmations.

\begin{figure}[tbp]
\centering
\includegraphics[width=0.92\linewidth]{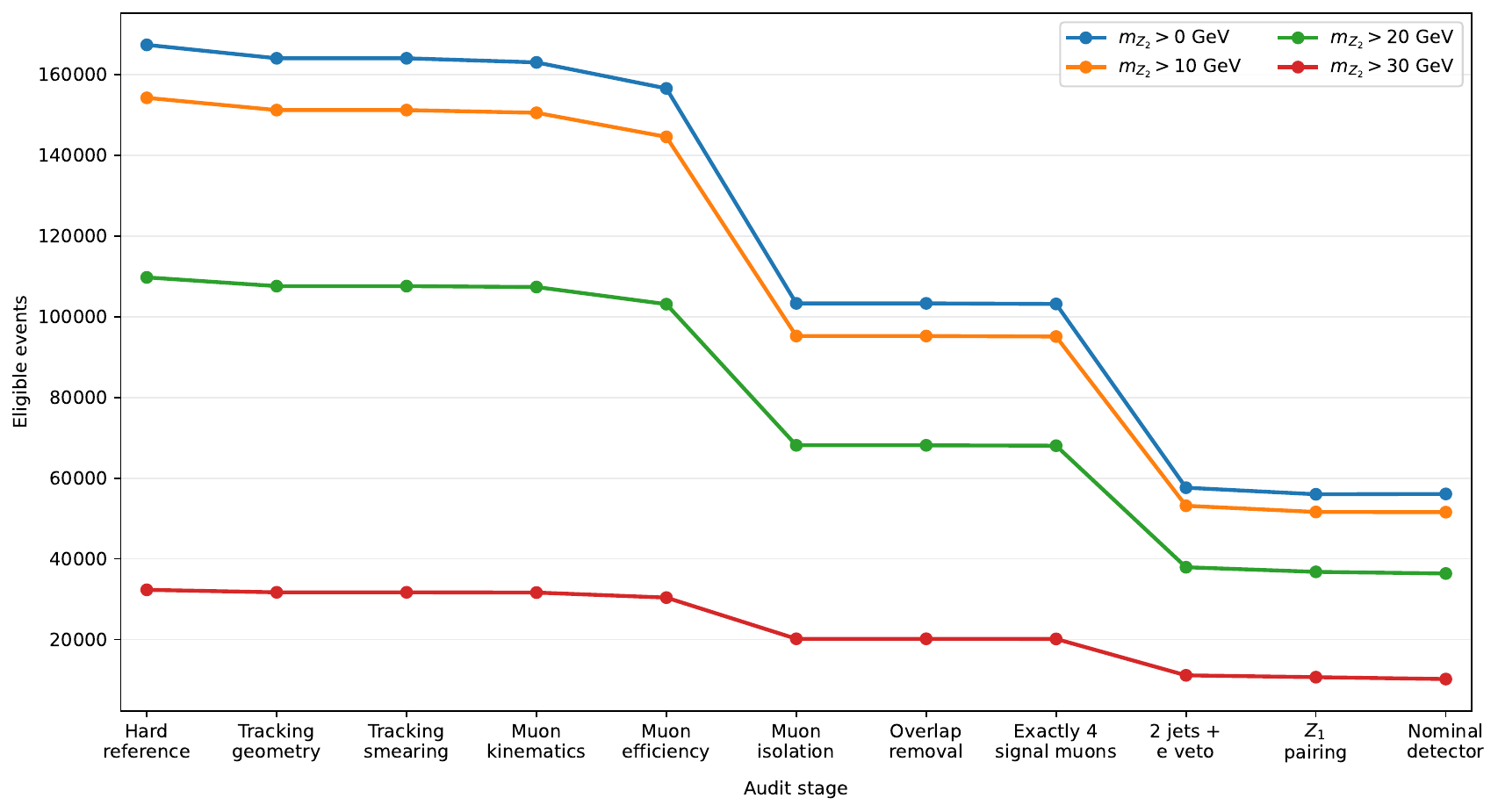}
\caption{
Eligible event counts through the frozen reconstruction and selection chain for the four nested $m_{Z_2}$ thresholds. The final detector-level counts are 56,091, 51,575, 36,375, and 10,224. The last stage includes kinematic migration effects and is therefore shown as an event count rather than interpreted as a standalone efficiency.
}
\label{fig:eventflow}
\end{figure}

The definitions in this section constitute the frozen HEP task used for the instruction-constrained benchmark. The next section describes how the agentic workflow executed this task and compares the resulting observable with the published reference only after the analysis configuration had been fixed.

\section{Agent execution and benchmark reproduction}
\label{sec:benchmark}

Having defined the agentic protocol in Sec.~\ref{sec:agentic} and the frozen HEP task in Sec.~\ref{sec:case}, we next evaluate whether the resulting workflow can execute the complete analysis and recover the published observable. This stage tests execution rather than open-ended discovery: the physics process, observable definition, and required analysis conventions are known, but the agent is instructed not to use the published numerical $I_3$ values as optimization targets. The formal comparison with Ref.~\cite{WuEtAl2025CEPC} is performed only after the analysis configuration has been fixed.

\subsection{Execution of the frozen analysis chain}

The agentic workflow implemented and connected the matrix-element generation, PYTHIA showering and hadronization, Delphes detector simulation, event extraction, reconstruction-stage bookkeeping, angular reconstruction, statistical evaluation, and plotting chain described in Sec.~\ref{sec:simulation}. Development proceeded through local and small-sample tests before the frozen configuration was propagated to the 200,000-event production. The software and physics issues encountered during this process---including the Delphes syntax incompatibility, the PYTHIA muon-mass convention, and HTCondor staging failures---were treated through the diagnostic pattern summarized in Table~\ref{tab:agentcases}, rather than by adjusting the analysis toward the published numerical result.

Once the estimator, reconstruction stages, selections, seed map, and bootstrap procedure had been fixed, the enlarged production was evaluated with deterministic scripts acting on the recorded artifacts. No detector parameter, estimator convention, or selection definition was subsequently changed to improve agreement with Ref.~\cite{WuEtAl2025CEPC}. The benchmark therefore tests whether the constructed workflow reaches the correct numerical regime and threshold dependence under an instruction-constrained reproduction, rather than whether an agent can tune an analysis to a visible target value.

\subsection{Quantitative comparison with the published observable}

Table~\ref{tab:benchmark} compares the final detector-level $I_3$ values obtained from the 200,000-event production with the corresponding numerical points reported in Ref.~\cite{WuEtAl2025CEPC}. The differences are $-0.049$, $-0.040$, $-0.059$, and $+0.042$ for the four nested $m_{Z_2}$ thresholds. Relative to the published central values, these correspond to differences of approximately $1.7\%$, $1.4\%$, $1.9\%$, and $1.4\%$, respectively. All four central values are therefore reproduced to within $2\%$, and the differences are within approximately one statistical standard deviation of the present sample.

\begin{table}[tbp]
\centering
\caption{Instruction-constrained benchmark comparison. Final detector-level $I_3$ values from the 200k production are shown together with the corresponding published CEPC values from Ref.~\cite{WuEtAl2025CEPC}. The formal comparison was performed after the present analysis configuration had been fixed; the published values were not used as tuning targets. The event counts are those of the present analysis. The four rows are nested selections and are therefore statistically correlated. $I_3^{\rm hard}$ is evaluated on the hard-reference population and $I_3^{\rm det}$ on the detector-level population given in the Events column.}
\label{tab:benchmark}
\begin{tabular}{rrrrr}
\toprule
$m_{Z_2}^{\min}$ [GeV] & Events & $I_3^{\rm hard}$ & $I_3^{\rm det}$ & Published $I_3$ \\
\midrule
0  & 56091 & 2.691 & 2.774 & 2.823 \\
10 & 51575 & 2.776 & 2.873 & 2.913 \\
20 & 36375 & 2.932 & 3.033 & 3.092 \\
30 & 10224 & 2.921 & 3.090 & 3.048 \\
\bottomrule
\end{tabular}
\end{table}

\begin{figure}[tbp]
\centering
\includegraphics[width=0.92\linewidth]{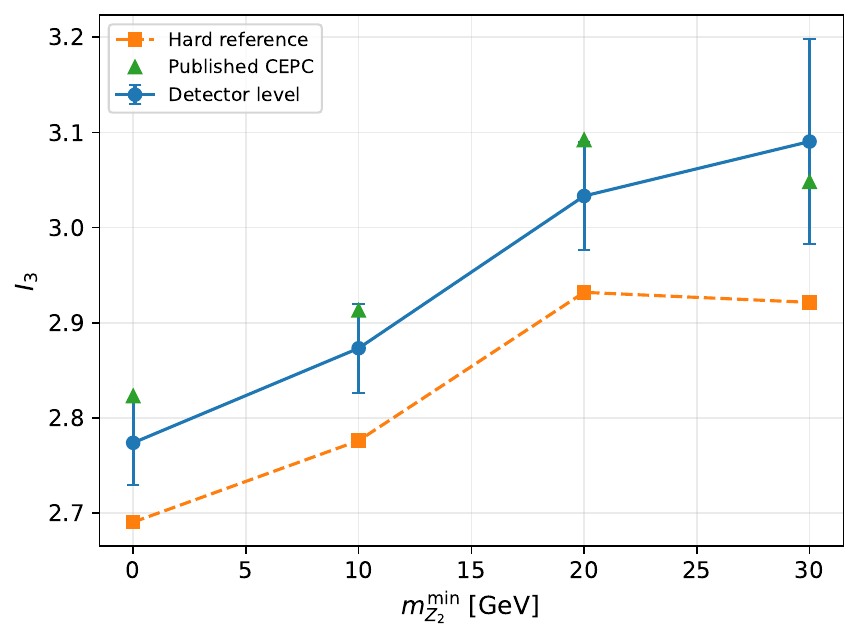}
\caption{
Instruction-constrained benchmark comparison of the hard-reference and detector-level
$I_3$ values as a function of the lower $m_{Z_2}$ threshold. Published values from
Ref.~\cite{WuEtAl2025CEPC} are shown for the formal post-freeze comparison and were
not used to tune the present analysis. Error bars on the detector-level points are
statistical bootstrap standard deviations.
}
\label{fig:finali3}
\end{figure}

The agreement applies specifically to the selected $I_3$ benchmark and should not be interpreted as demonstrating that every observable or implementation detail in Ref.~\cite{WuEtAl2025CEPC} has been reproduced to the same precision. Its role in this study is instead to provide an external quantitative validation that the agent-constructed workflow reproduces the scale and $m_{Z_2}$ dependence of the published observable without post-comparison retuning.

\subsection{From reproduction to an unexplained physics shift}

The benchmark comparison establishes that the detector-level calculation is numerically consistent with the published observable, but it also exposes a second feature of the present analysis: the final detector-level $I_3$ values are systematically displaced from the corresponding hard-reference values. This hard-to-detector shift was not itself the target of the reproduction and therefore provides the starting point for the subsequent open-ended investigation.

The total shift is evaluated with 5,000 paired-bootstrap~\cite{Efron1979} replicas. For each replica, the same resampled event multiplicities are applied simultaneously to the hard-reference and detector-level estimators. This preserves their event-level covariance and provides the uncertainty on the difference directly, rather than treating the two endpoints as statistically independent samples.

Table~\ref{tab:totalshift} gives the resulting hard-to-detector shifts and percentile 95\% intervals. The first three intervals are positive in this sample, whereas the interval for $m_{Z_2}>30$~GeV includes zero. Because the four thresholds correspond to nested event populations, the results are strongly correlated and are not interpreted as four independent confirmations.

\begin{table}[tbp]
\centering
\caption{Paired-bootstrap hard-to-detector shifts in the present 200k analysis. Uncertainties are bootstrap standard deviations and the intervals are percentile 95\% intervals. The four selections are nested and therefore strongly correlated.}
\label{tab:totalshift}
\begin{tabular}{rrrr}
\toprule
$m_{Z_2}^{\min}$ [GeV] & $\Delta I_3$ & $\sigma_{\rm paired}$ & 95\% interval \\
\midrule
0  & $+0.083$ & 0.036 & $[0.012,\,0.153]$ \\
10 & $+0.097$ & 0.038 & $[0.021,\,0.170]$ \\
20 & $+0.101$ & 0.046 & $[0.008,\,0.189]$ \\
30 & $+0.169$ & 0.088 & $[-0.006,\,0.340]$ \\
\bottomrule
\end{tabular}
\end{table}

\begin{figure}[tbp]
\centering
\includegraphics[width=0.92\linewidth]{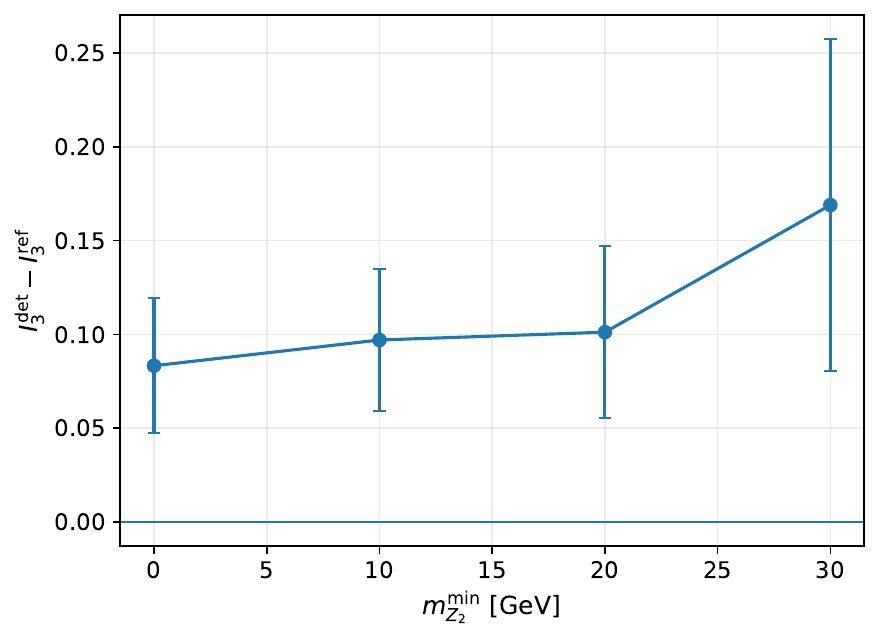}
\caption{
Paired hard-to-detector shift in $I_3$ for the four nested $m_{Z_2}$ selections. Error bars represent paired-bootstrap standard deviations. The selections are strongly correlated and are not interpreted as independent tests.
}
\label{fig:totalshift}
\end{figure}
This stage therefore provides two distinct results for the scientific-agent evaluation. First, the agent-constructed analysis reproduces the selected published $I_3$ central values to within $2\%$ across all four thresholds under the instruction-constrained protocol. Second, after the benchmark is established, the frozen analysis contains a statistically resolved hard-to-detector shift whose origin is not specified by the reference result. The next task is therefore qualitatively different from reproduction: using the frozen production and estimator, the agentic workflow is asked to determine which reconstruction stage generates the shift and whether a physically interpretable mechanism can be identified without a predefined target answer.

\section{Open-ended agentic diagnosis of the reconstruction chain}
\label{sec:sculpting}

Having established the instruction-constrained benchmark, the analysis moves to a qualitatively different task: open-ended diagnosis of an unexplained physics shift. No target reconstruction stage, expected numerical contribution, or preferred physical explanation was supplied. Using the frozen production and estimator, the agentic workflow was asked to determine which step in the reconstruction and selection chain is primarily responsible for the observed hard-to-detector change in $I_3$.

The diagnostic strategy was to decompose the detector chain into explicit reconstruction stages and to construct complementary audits that separate population selection from reconstructed-vector response. This step was not predefined by a published target result; it emerged from the need to explain the benchmark-stage discrepancy between the hard-reference and detector-level estimators.

\subsection{Fixed-vector mask audit}

The first diagnostic constructed by the agentic workflow holds the hard-reference angular kernels fixed while successively applying the detector-stage selection masks. This deliberately isolates a narrow question: how much does $I_3$ change purely because different subsets of the same underlying events survive each reconstruction step?

Because the four-vectors are immutable in this construction, a stage that only modifies momenta is expected to produce zero direct shift unless that modification changes a later selection mask. Such zeros are therefore a consequence of the diagnostic definition rather than evidence that the physical detector has no momentum response.

For the inclusive $m_{Z_2}>0$ selection, the geometry, kinematic, and stochastic-efficiency stages each change $I_3$ by at most a few $\times10^{-3}$. The largest single transition occurs when muon isolation is applied:
\begin{equation}
 \Delta I_3^{\rm iso}=0.043\pm0.019.
\end{equation}
The same qualitative pattern persists at the higher $m_{Z_2}$ thresholds, as summarized in Table~\ref{tab:maskvector}. Subsequent exactly-four-muon, jet/electron-veto, pairing, and final-stage migration effects are retained explicitly in the audit rather than absorbed into a single detector correction.

\begin{figure}[tbp]
\centering
\includegraphics[width=0.92\linewidth]{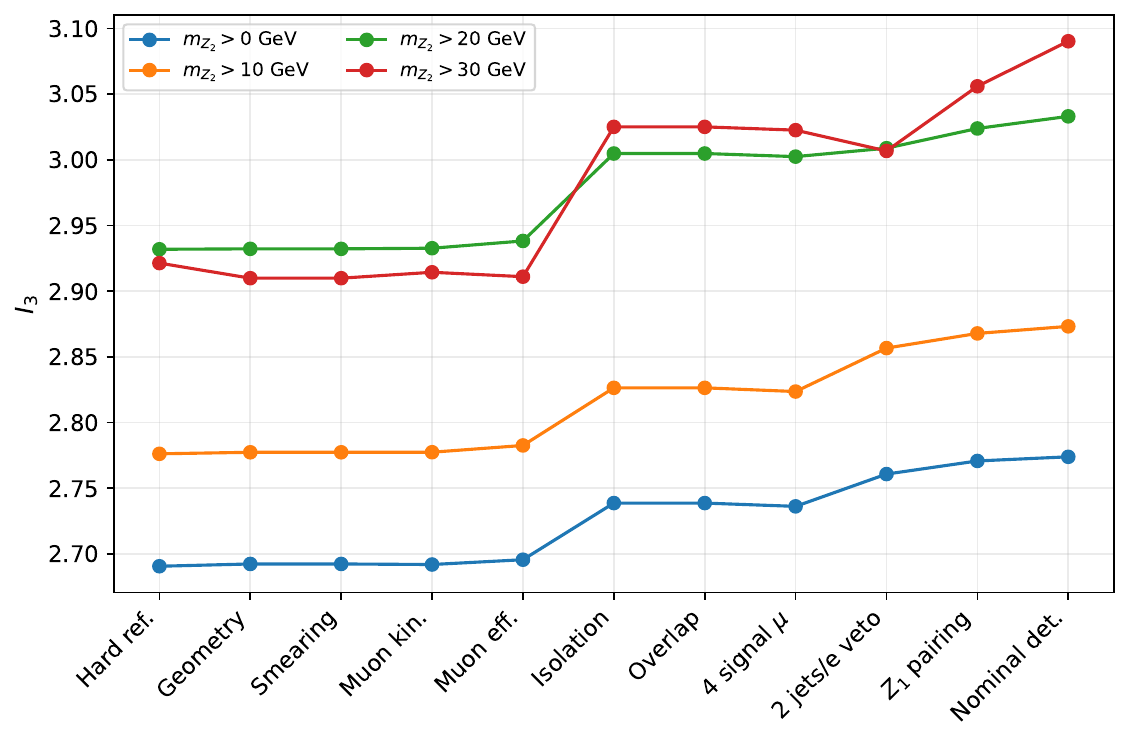}
\caption{
Stage-resolved $I_3$ in the fixed-vector mask audit. For all stages except the last, the hard-reference angular kernels are held fixed and only the accepted event population changes, so motion along a curve represents selection-induced sculpting. The final nominal-detector point recomputes the pairing and decay angles from reconstructed muons and equals $I_3^{\rm det}$ in Table~\ref{tab:benchmark}.
}
\label{fig:stagei3}
\end{figure}

The fixed-vector audit therefore identifies muon isolation as the dominant single selection stage in the hard-to-detector change. However, because this construction intentionally freezes the hard-reference vectors, an additional control is required to test whether the conclusion survives when reconstructed kinematics are propagated.

\subsection{Complementary stage-vector audit}

The agentic workflow therefore constructed a second audit in which each reconstruction stage uses both its corresponding selection mask and its stage-specific reconstructed four-vectors. This stage-vector construction includes both population changes and the kinematic response accumulated up to that point.

The purpose is not to subtract the two audits as statistically independent measurements. Instead, their agreement provides a robustness test: if the isolation-stage effect were created mainly by the artificial freezing of the hard-reference vectors, the corresponding stage-vector shift would be expected to differ substantially.

\begin{table}[tbp]
\centering
\caption{Isolation-stage shift in the fixed-vector mask audit and the complementary stage-vector audit. Uncertainties are paired-bootstrap standard deviations.}
\label{tab:maskvector}
\begin{tabular}{rrr}
\toprule
$m_{Z_2}^{\min}$ [GeV] & Fixed-vector $\Delta I_3$ & Stage-vector $\Delta I_3$ \\
\midrule
0  & $0.043\pm0.019$ & $0.040\pm0.019$ \\
10 & $0.044\pm0.020$ & $0.039\pm0.020$ \\
20 & $0.067\pm0.024$ & $0.065\pm0.023$ \\
30 & $0.114\pm0.044$ & $0.126\pm0.043$ \\
\bottomrule
\end{tabular}
\end{table}

\begin{figure}[tbp]
\centering
\includegraphics[width=0.92\linewidth]{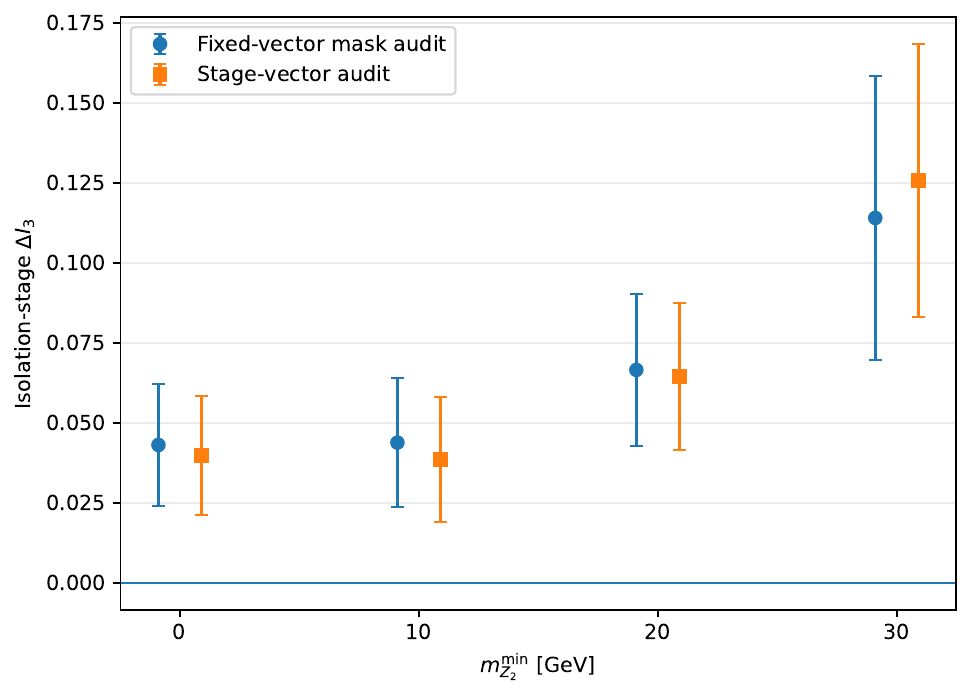}
\caption{
Isolation-stage shift from the fixed-vector mask audit compared with the stage-vector audit. The agreement between the two constructions shows that the identified isolation effect persists when stage-specific reconstructed four-vectors are propagated. Error bars are paired-bootstrap standard deviations; points are offset horizontally for visibility.
}
\label{fig:maskvector}
\end{figure}

The two constructions give consistent isolation-stage shifts within uncertainties at all four thresholds. The dominant isolation contribution is therefore not an artifact of freezing the hard-reference four-vectors. In the terminology used here, \emph{selection sculpting} refers specifically to the fixed-vector mask result, while \emph{stage-vector response} includes the reconstructed kinematics available at each stage. The total hard-to-detector difference contains both effects as well as shower and migration contributions and is not interpreted as pure detector smearing.

For the scientific-agent evaluation, this stage marks the transition from reproduction to diagnosis. The workflow was not given the answer that isolation should dominate; it arrived at that conclusion by decomposing the reconstruction chain, constructing a diagnostic that isolates acceptance effects, and then testing the result with an independent kinematic treatment. Identifying the dominant stage still does not explain why the Bell observable is changed, however. The next section therefore continues the open-ended investigation by asking whether the isolation inefficiency can be connected to the tomographic angles and to an upstream physical feature of the event topology.

\section{Agent-led physical investigation of the isolation effect}
\label{sec:mechanism}

The stage-resolved diagnosis identifies muon isolation as the dominant single reconstruction stage affecting $I_3$, but this does not yet explain why the Bell observable is changed. The next task is therefore open-ended: using the same frozen production, the agentic workflow is asked to determine whether the isolation inefficiency is structured in the tomographic variables and, if so, whether that structure can be traced to an upstream feature of the event and to a physically interpretable source of cone activity. No target angular dependence, topology variable, or expected composition of the isolation cone was supplied in advance.

\subsection{Does isolation depend on the tomographic angles?}
\label{sec:angeff}

The first hypothesis tested is that isolation may not remove events uniformly in the variables entering the tomography. If isolation acted as a random loss independent of the decay angles, it would primarily reduce the event count rather than reshape the density-matrix estimator. The agentic workflow therefore constructs the conditional isolation efficiency in bins of the same angular variables used to evaluate the Bell observable. The efficiency is defined per signal muon as the fraction of pre-isolation signal muons passing isolation in events for which the pairing and angular variables are defined. Statistical intervals are obtained by resampling event clusters so that correlations among muons from the same event are retained.

The dependence on $\cos\theta_1$ and $\cos\theta_2$ is mild, with efficiencies varying by only a few per mille across the accepted range. The azimuthal variables show a clearer structure: the efficiencies span approximately 0.887--0.901 in both $\phi_1$ and $\phi_2$, corresponding to a modulation of about 1.4 percentage points, substantially larger than the per-bin bootstrap uncertainty of roughly $10^{-3}$. The isolation requirement therefore acts as an angle-dependent acceptance function rather than as a uniform event-loss factor.

\begin{figure}[tbp]
\centering
\includegraphics[width=0.92\linewidth]{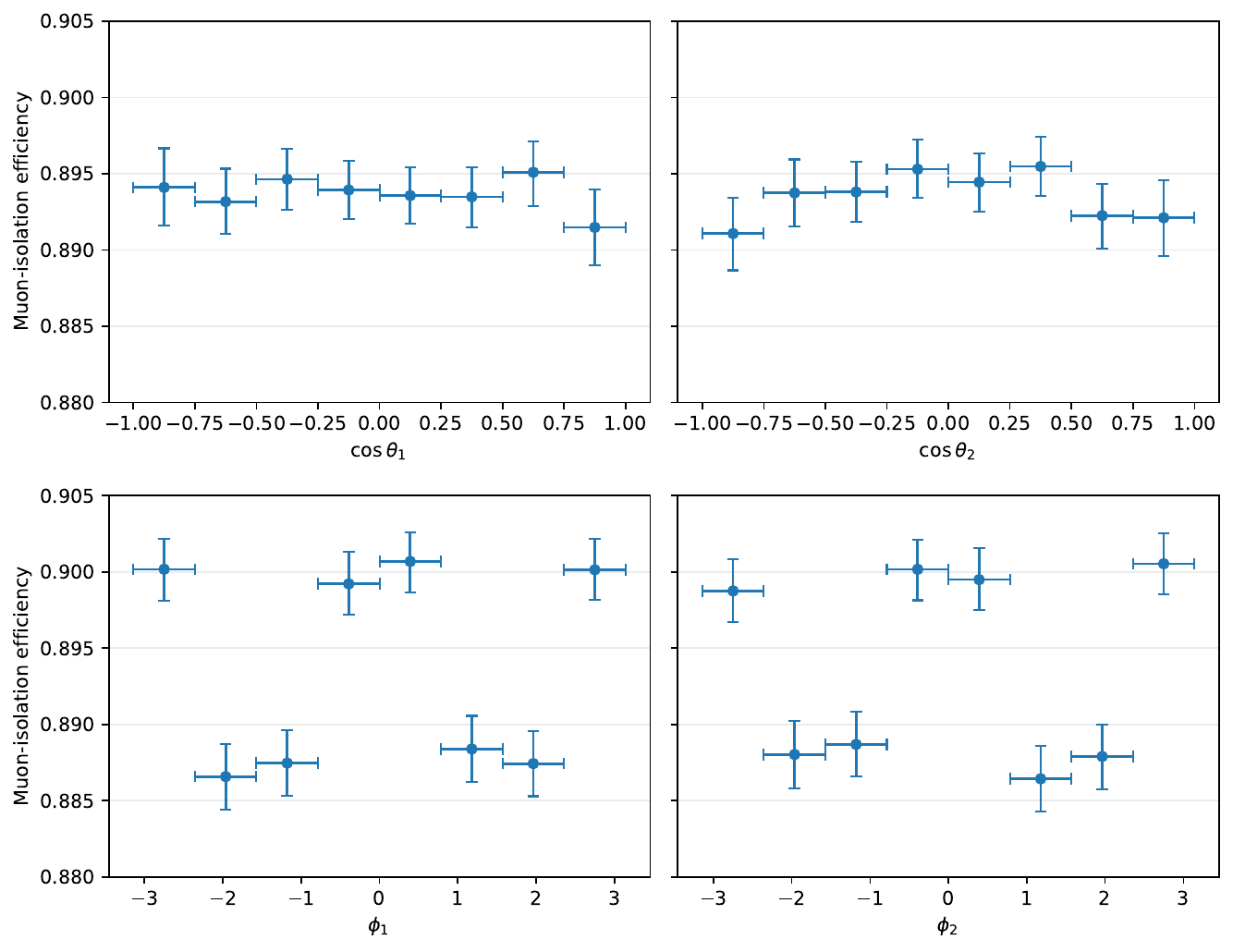}
\caption{
Muon-isolation efficiency versus the four angular variables entering the tomography. The efficiency is defined per signal muon; vertical intervals are bootstrap 95\% intervals with event-level resampling, and horizontal bars indicate the bin widths. The polar-angle dependence is mild, while a clear azimuthal modulation is present.
}
\label{fig:angulareff}
\end{figure}

This establishes that the isolation stage can directly reweight the variables entering the Bell estimator. The next question is whether the observed angular non-uniformity is associated with a physically meaningful feature of the event topology.

\subsection{Which upstream topology predicts isolation failure?}
\label{sec:hardquark}

The first candidate considered was reconstructed-jet proximity, because nearby hadronic activity provides an intuitive explanation for isolation failure. The agentic workflow explicitly tested this interpretation rather than adopting it directly. The control revealed that the pre-overlap jet collection is not a suitable upstream probe because it can contain the signal muon itself, while the final cleaned-jet collection is defined downstream of the relevant isolation decision. As discussed in Sec.~\ref{sec:agentic} and Appendix~\ref{app:controls}, these reconstructed-jet proxies were therefore rejected.

The diagnostic sequence then moved to a truth-level upstream variable that is independent of detector object cleaning: the two status-23 hard quarks from the uniquely identified associated $Z\to q\bar q$ decay. The hadronic $Z$ is uniquely identified in all 200,000 events, with no missing or ambiguous cases. For each signal muon we define
\begin{equation}
 \Delta R_{\mu q}^{\min}
 =
 \min\left[
 \Delta R(\mu,q),
 \Delta R(\mu,\bar q)
 \right].
\end{equation}

A strong dependence emerges. The isolation efficiency is only $8.6\%$ for $\Delta R_{\mu q}^{\min}<0.25$ and $14.3\%$ for $0.25<\Delta R_{\mu q}^{\min}<0.50$. It rises to $67.7\%$ in the interval 0.50--0.75, to $90.7\%$ in 0.75--1.00, and reaches a $97$--$98\%$ plateau for well-separated muons. The failed-muon sample has a mean hard-quark distance of 0.767, compared with 1.726 for passing muons. This links the isolation inefficiency to the upstream hadronic topology without relying on a detector object defined downstream of the isolation decision.

\begin{table}[tbp]
\centering
\caption{Muon-isolation efficiency versus the minimum distance to the two hard quarks from the associated $Z\to q\bar q$ decay. Uncertainties are bootstrap standard deviations.}
\label{tab:hardquarkeff}
\begin{tabular}{rrrr}
\toprule
$\Delta R_{\mu q}^{\min}$ & Passing & Total & Efficiency \\
\midrule
$[0.00,0.25)$ & 878 & 10182 & $0.0862\pm0.0029$ \\
$[0.25,0.50)$ & 4380 & 30651 & $0.1429\pm0.0020$ \\
$[0.50,0.75)$ & 34259 & 50587 & $0.6772\pm0.0021$ \\
$[0.75,1.00)$ & 63355 & 69836 & $0.9072\pm0.0011$ \\
$[1.00,1.25)$ & 81178 & 85642 & $0.9479\pm0.0008$ \\
$[1.25,1.50)$ & 91588 & 94899 & $0.9651\pm0.0006$ \\
$[1.50,1.75)$ & 97618 & 100278 & $0.9735\pm0.0005$ \\
$[1.75,2.00)$ & 97320 & 99738 & $0.9758\pm0.0005$ \\
$[2.00,2.25)$ & 85994 & 88076 & $0.9764\pm0.0005$ \\
$[2.25,2.50)$ & 61914 & 63360 & $0.9772\pm0.0006$ \\
$[2.50,3.00)$ & 71915 & 73524 & $0.9781\pm0.0006$ \\
$[3.00,3.50)$ & 18499 & 18870 & $0.9803\pm0.0011$ \\
$[3.50,4.50)$ & 1207 & 1221 & $0.9885\pm0.0032$ \\
\midrule
All & 710105 & 786864 & $0.9024$ \\
\bottomrule
\end{tabular}
\end{table}

\begin{figure}[tbp]
\centering
\includegraphics[width=0.92\linewidth]{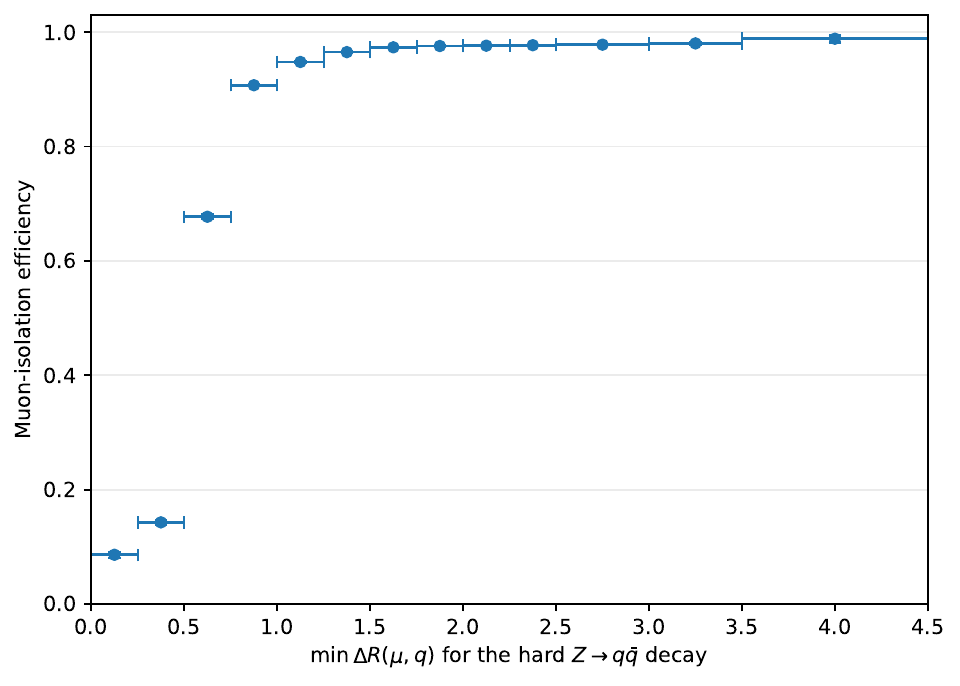}
\caption{
Muon-isolation efficiency as a function of the minimum distance to either hard quark from the associated $Z\to q\bar q$ decay. This upstream truth variable is independent of detector jet cleaning. Error bars are obtained from bootstrap resampling of the 200 independent 1000-event generation blocks; horizontal bars indicate the bin widths.
}
\label{fig:hardquark}
\end{figure}

The strong dependence on $\Delta R_{\mu q}^{\min}$ indicates that proximity to the hard hadronic system is a useful predictor of isolation failure. The remaining question is what physical activity inside the isolation cone accompanies this topology.

\subsection{What produces the electromagnetic cone activity?}

The final diagnostic step investigates the electromagnetic component of the isolation cone. The cone also contains charged- and neutral-hadron contributions, but the photon component can be traced directly through the generator ancestry. Stable photons with $p_T>0.5$~GeV and $\Delta R<0.5$ around each pre-isolation signal muon are classified as muon-FSR-like if any muon occurs in the ancestry; if no muon ancestor is present, a hadronic ancestor defines the hadron-decay category; the remaining photons are classified as direct parton-shower or other contributions.

In the full 200k sample, $15.0\%$ of signal-muon candidates contain at least one such stable truth photon. The scalar truth-photon $p_T$ is divided into $43.2\%$ muon-FSR-like, $56.2\%$ hadron-decay, and $0.63\%$ direct parton-shower contributions. The truth-photon scalar-$p_T$ proxy is strongly correlated with the electromagnetic isolation activity stored by Delphes, with Pearson correlation coefficient $\rho=0.9863$.

The pass/fail comparison supports the mixed-origin interpretation. For passing muons, the mean FSR-like and hadron-decay photon sums are 0.105 and 0.053~GeV, respectively. For failing muons, they rise to 2.74 and 4.34~GeV. The corresponding mean stored electromagnetic cone activity is 0.161~GeV for passing muons and 7.60~GeV for failing muons. The photon component of the cone activity therefore cannot be attributed to QED final-state radiation or hadron decays alone; both contribute, with a somewhat larger hadron-decay share. The charged- and neutral-hadron components of the isolation sum are not decomposed here.

\begin{table}[tbp]
\centering
\small
\caption{Generator-ancestry decomposition of the scalar truth-photon $p_T$ within $\Delta R<0.5$ of the pre-isolation signal muon.}
\label{tab:photonorigin}
\begin{tabularx}{\linewidth}{
>{\raggedright\arraybackslash}X
>{\centering\arraybackslash}p{0.22\linewidth}
>{\centering\arraybackslash}p{0.22\linewidth}
>{\centering\arraybackslash}p{0.22\linewidth}}
\toprule
Origin &
Fraction of truth-photon $p_T$ &
Mean per passing muon [GeV] &
Mean per failing muon [GeV] \\
\midrule
Muon-FSR-like & 0.432 & 0.105 & 2.741 \\
Hadron decay  & 0.562 & 0.053 & 4.337 \\
Parton shower & 0.006 & 0.001 & 0.048 \\
\bottomrule
\end{tabularx}
\end{table}

The open-ended investigation therefore forms a sequential diagnostic chain rather than a collection of predetermined physics plots. The agentic workflow first establishes that isolation is angle dependent, then rejects an initially plausible but non-causal reconstructed-jet proxy, identifies the hard $Z\to q\bar q$ system as an upstream topology reference, and finally traces the electromagnetic cone activity to a mixture of muon final-state radiation and hadron decays. None of these intermediate targets was supplied as a predefined answer. This sequence provides the strongest evidence in the present case study that the workflow moved beyond mechanical reproduction and participated in an open-ended, physically interpretable HEP investigation.

\section{Discussion}
\label{sec:discussion}

The main purpose of this study is not to establish a new strategy for Bell-inequality measurements at the CEPC, but to use a realistic collider-phenomenology analysis as a controlled test of how an LLM-based scientific agent can participate in a multistep research workflow. The case study probes two complementary capabilities. First, the agent carries out an instruction-constrained reconstruction of a previously published observable: the physics objective, required analysis steps, and conventions are supplied, while the agent is explicitly instructed not to use the published numerical values as tuning targets. No technical information-access barrier is imposed, so this should not be interpreted as a formally blinded benchmark. Second, after the benchmark analysis is fixed and compared with the published result, the agent is asked to investigate an open-ended question for which no target answer is supplied. In that stage, the dominant reconstruction effect, the relevant topology variable, and the physical interpretation of the isolation dependence emerge through successive diagnostic controls rather than from a predefined solution. Taken together, these two stages test not only whether an agent can execute an established analysis, but also whether it can support the iterative reasoning required to extend that analysis.

The case study suggests that reliability does not come from unrestricted autonomy, but from separating agentic exploration from the evidential standard required for a physics result. The agent was useful for integrating heterogeneous information---source code, detector cards, logs, intermediate files, event-level outputs, and statistical results---and for moving rapidly from an observed inconsistency to a proposed explanation and a targeted control. At the same time, numerical conclusions entered the analysis only through deterministic scripts acting on pinned inputs. Frozen definitions, fixed random seeds, checksums, isolated diagnostic branches, paired statistical tests, and explicit human authorization prevented exploratory reasoning from silently changing the nominal result. The 1000-event isolation anomaly provides a concrete example: an initially striking effect was not promoted directly to a conclusion, but instead triggered additional controls and an enlarged frozen-definition analysis, which produced a substantially smaller and statistically quantified shift. Human participation therefore remains essential, but its role changes: the researcher defines the scientific objective, approves state-changing actions and changes to frozen definitions, evaluates the physical meaning of proposed controls, and retains responsibility for the final interpretation, while the agent performs much of the labor-intensive intermediate work of inspection, implementation, debugging, control construction, and consistency checking.

This division of labor is particularly relevant to collider phenomenology and experimental analysis, where a large fraction of research time is spent not on the final physics inference itself but on maintaining long software chains, tracing configuration details, reproducing selections, debugging production failures, validating intermediate objects, and repeating related diagnostic studies. The potential advantage of scientific agents is therefore not simply faster code generation, but faster scientific iteration. A more meaningful productivity metric is how many validated scientific questions can be advanced per unit of human researcher time. Earlier HEP-oriented work has already shown that retrieval-augmented and agent-assisted systems can reduce the effort required to locate, integrate, and verify domain-specific scientific evidence in muon-collider studies and silicon-detector R\&D~\cite{Jiang2026MuonRAG,Gao2026SiliconRAG}; the present study extends this direction from evidence access toward the execution and diagnosis of a multistep collider analysis. An agent that can maintain analysis state across multiple tools, regenerate samples, compare configurations, propagate changes through a workflow, construct validation plots, and preserve provenance could allow one researcher to supervise several computational investigations that would otherwise have to be pursued sequentially. For LHC analyses, the same pattern could extend to substantially larger software environments in which detector reconstruction, object selections, systematic variations, batch production, statistical models, and documentation must remain synchronized. The present work does not prospectively measure active human time and therefore does not support a numerical claim for a speed-up factor, but it motivates future benchmarks that record both total computational time and direct human intervention time.

There are, however, important limits to the conclusion. The present study uses a single physics channel, fast detector simulation, and explicit human supervision. It does not demonstrate that an agent can independently certify a complete experimental analysis, choose all physically relevant systematic uncertainties, or replace collaboration review. Successful reconstruction of one observable also does not establish robustness across arbitrary collider tasks. A broader evaluation should therefore move from isolated demonstrations to systematic benchmarking across a large set of historical HEP analyses. Published studies provide a particularly useful test bed because the scientific questions and final answers already exist, while selected information and numerical targets can be hidden or restricted during execution. Such a benchmark could measure numerical accuracy, failure recovery, number and type of human interventions, computational cost, active researcher time, and the fraction of tasks completed without escalation.

The CEPC Bell-observable analysis should therefore be viewed as a proof of concept for a broader research program rather than as evidence that autonomous HEP analysis is already solved. The most important result of the present case study is that the agentic workflow moved through two distinct regimes: instruction-constrained reconstruction of a known calculation and open-ended diagnosis of a previously unspecified detector effect. This suggests a practical path toward more general collider-phenomenology agents whose value is measured not by autonomy alone, but by whether they can reproducibly expand the number and complexity of analyses that a physicist, PhD student, postdoctoral researcher, or small group can pursue within a fixed human-time budget.

\section{Conclusion}
\label{sec:conclusion}

This study uses a realistic collider-phenomenology analysis as a controlled test of how far an LLM-based scientific agent can participate in computational particle-physics research. The evaluation probes two different capabilities. First, the agent carries out an instruction-constrained reconstruction of a previously published CEPC Bell-observable analysis: the physics objective, required conventions, and analysis procedure are available, while the agent is explicitly instructed not to use the published numerical results as tuning targets. No technical information-access barrier is imposed, and the exercise is therefore not claimed to be formally blinded. The resulting detector-level $I_3$ central values reproduce the four published benchmark points to within approximately $2\%$, with no post-comparison retuning of detector parameters or analysis selections. Second, after establishing this benchmark, the agent is asked to pursue an open-ended extension for which the dominant reconstruction effect, the relevant topology variable, and the physical explanation are not supplied in advance.

The CEPC case study provides a concrete demonstration of this second capability. Using 200,000 fast-simulated events and a stage-resolved audit, the analysis identifies muon isolation as the largest single contribution to the hard-to-detector shift of the CGLMP estimator $I_3$. For the inclusive selection, the isolation-stage shift is $0.043\pm0.019$ in the fixed-vector construction and $0.040\pm0.019$ in the stage-vector construction. The subsequent diagnostic sequence shows that the isolation efficiency is non-uniform in the tomographic angles and strongly dependent on the distance to the hard quarks from the associated $Z\to q\bar q$ decay, rising from $8.6\%$ for $\Delta R^{\min}_{\mu q}<0.25$ to a $97$--$98\%$ plateau for well-separated muons. Generator-ancestry tracing further shows that the electromagnetic cone activity has mixed origin, with $43.2\%$ of the scalar truth-photon $p_T$ classified as muon-FSR-like and $56.2\%$ associated with hadron decays. These physics results are not the primary motivation of the study, but they show that the agentic workflow can move beyond mechanical reproduction and participate in a sequential, physically interpretable investigation.

The main methodological conclusion is that scientific agents are most useful when exploratory reasoning is separated from the evidential chain used for the final result. In the present workflow, the agent participates in workflow design, sample production, configuration inspection, anomaly diagnosis, control construction, statistical analysis, and result organization, while frozen definitions, fixed random seeds, checksum-pinned artifacts, isolated diagnostic branches, deterministic scripts, paired statistical tests, and explicit human authorization preserve provenance and control over the final interpretation. The agent therefore acts not as a replacement for the physicist, but as a research collaborator that can reduce technical overhead and shorten the cycle between a scientific question, a computational test, and a validated answer. The potentially important productivity gain is not simply faster code generation, but the ability to advance more validated scientific questions within a fixed human-time budget.

A natural next step is to move from this single case study to a systematic benchmark
of scientific-agent capability across a large collection of historical HEP analyses.
Published studies provide useful evaluation tasks because the scientific questions
and final results already exist, while numerical targets and selected information
can be restricted during execution. Such a program could quantify numerical accuracy,
robustness to software and physics ambiguities, failure recovery, computational cost,
the number and type of human interventions, and active researcher time. Multi-agent
architectures may provide one practical route to scaling such studies, with specialized
agents for literature interpretation, event generation, detector simulation, analysis
construction, statistical validation, systematic checks, and critical review working
in parallel under shared provenance constraints. The broader objective, however, is
not autonomy for its own sake, but to determine whether AI agents can reproducibly
increase the number and complexity of HEP analyses that an individual researcher or
small group can pursue within a fixed human-time budget.
\section*{Data and software availability}

The analysis code, configuration files, workflow definitions, validation tests, and scripts used to reproduce the results presented in this work are available at
\url{https://github.com/AItutorialjrb/agentic_HZZ_case}.

The frozen detector-level ROOT files, derived analysis products, publication tables, and provenance metadata associated with the 200,000-event production are archived at
\url{https://huggingface.co/datasets/Ruobingjiang/AI-Agent-HEP-HZZ-Case-Study}.

Together, these repositories provide the software, configuration, frozen data products, random-seed information, and machine-readable outputs required to reproduce the main numerical results and publication figures.



\FloatBarrier
\appendix
\section{Development-sample and convention controls}
\label{app:controls}

\subsection{Why the 1000-event isolation shift was not used}

The initial single-seed diagnostic used an earlier population definition and produced an isolation-stage shift of approximately $+0.661$.  Applying the final stage definition to the same seed gives the paired shifts in Table~\ref{tab:devsample}.

\begin{table}[!htbp]
\centering
\caption{Isolation-stage $\Delta I_3$ in the 1000-event development sample, re-evaluated with the final stage definitions.  These values are not used for any conclusion.}
\label{tab:devsample}
\begin{tabular}{rrrr}
\toprule
$m_{Z_2}^{\min}$ [GeV] & $\Delta I_3$ & Bootstrap std. & 95\% interval \\
\midrule
0  & $+0.235$ & 0.273 & $[-0.310,\,0.767]$ \\
10 & $+0.219$ & 0.289 & $[-0.345,\,0.803]$ \\
20 & $+0.347$ & 0.334 & $[-0.302,\,1.012]$ \\
30 & $+1.457$ & 0.571 & $[0.411,\,2.612]$ \\
\bottomrule
\end{tabular}
\end{table}
The high-threshold line contains only 93 post-isolation events and is a useful illustration of the instability of the small development sample rather than an independent physics result.  The final 200k analysis is therefore the only quantitative basis for the conclusions in the main text.

\subsection{Muon-mass import and FSR convention}

The generated LHE sample serializes the four final-state muons with zero mass.  PYTHIA's default mapping to physical lepton masses can redistribute recoil among the imported hard particles even when the total four-muon system is nearly conserved.  In a single-seed control this produced a maximum $m_{Z_2}$ change of about 0.448 GeV, substantially larger than the muon rest mass because it is a recoil effect rather than a direct scalar mass addition.  The final workflow records both the original LHE hard vectors and the showered event, allowing this convention dependence to be audited explicitly.  The fixed hard-reference estimator always uses the original LHE vectors.

\subsection{Rejected jet-proximity definitions}

A pre-overlap reconstructed jet collection was initially considered as a proxy for the hadronic environment.  It was rejected after 37.5\% of matched signal muons were found to have a nearest jet at exactly $\Delta R=0$, demonstrating that the collection contains self-overlapping muon-associated jets.  The final cleaned-jet collection removes most of these objects but is downstream of object-cleaning decisions and is therefore not causal with respect to the isolation stage.  Inclusive generator jets also mix signal-muon/FSR-related clustering with the hadronic system.  The main text therefore uses only the two hard quarks from the uniquely identified associated $Z\to q\bar q$ decay.

\section{Reproducibility record}
\label{app:repro}

The production and audit are organized around immutable event identities $(\mathrm{seed},\mathrm{event\ index})$.  The 200k sample comprises 200 seeds with 1000 events each.  The final extraction validated all 200,000 identities with no duplicates, missing events, or partial extracts.  The diagnostic matching between extracted signal muons and the Delphes pre-isolation collection was validated for all 786,864 stored signal-muon candidates with no failed events; the maximum absolute matching differences in $p_T$ and $\eta$ were at numerical roundoff level ($\mathcal{O}(10^{-14})$).

The main paired-bootstrap estimator resamples underlying event identities with the same multiplicity vector applied simultaneously to all compared stages.  This preserves the covariance that would be lost by treating the hard and detector estimators as independent samples.  Mechanism plots involving per-muon quantities preserve within-event correlations by clustering the resampling unit at event or independent generation-block level, as stated in the corresponding captions.

The repository records exact random seeds, software versions, resolved steering, file checksums, Git revision, command line, host, and timestamps for production artifacts.  Diagnostic outputs are additive and do not overwrite frozen production.  The publication package is designed so that all figures can be regenerated from machine-readable CSV/JSON tables rather than extracted from the manuscript PDF.

\bibliographystyle{unsrt}
\bibliography{refs}

\end{document}